\documentclass[conference]{IEEEtran}
\IEEEoverridecommandlockouts
\usepackage{cite}
\usepackage{amsmath,amssymb,amsfonts}
\usepackage{algorithmic}
\usepackage{graphicx}
\usepackage{textcomp}
\usepackage{xcolor}

\usepackage{minted}
\usepackage{listings}
\usepackage{multirow}
\usepackage{multicol}
\usepackage{booktabs}
\usepackage{hyperref}

\ifCLASSOPTIONcompsoc
  \usepackage[caption=false,font=normalsize,labelfont=sf,textfont=sf]{subfig}
\else
  \usepackage[caption=false,font=footnotesize]{subfig}
\fi

\def\BibTeX{{\rm B\kern-.05em{\sc i\kern-.025em b}\kern-.08em
    T\kern-.1667em\lower.7ex\hbox{E}\kern-.125emX}}

\usepackage{fancyhdr}

\fancypagestyle{firstpage}{
  \fancyhf{} 
  
  \fancyhead[C]{\small 2025 34th International Conference on Parallel Architectures and Compilation Techniques (PACT)}
  
  \fancyfoot[C]{\begin{minipage}{\textwidth}\centering \scriptsize
    \copyright 2025 IEEE. Personal use of this material is permitted. Permission from IEEE must be obtained for all other uses, in any current or future media, including reprinting/republishing this material for advertising or promotional purposes, creating new collective works, for resale or redistribution to servers or lists, or reuse of any copyrighted component of this work in other works. \\
    \vspace{2mm}
    DOI: \href{https://doi.org/10.1109/PACT65351.2025.00028}{10.1109/PACT65351.2025.00028}
  \end{minipage}}
}

\begin{document}

\title{LOOPer: A Learned Automatic Code Optimizer For Polyhedral Compilers
\thanks{Work was conducted while the authors were at their respective institutions as listed. The affiliations of A. Boudaoud, N. Tchoulak, H. Benyamina, and H. Leather have since changed.}}

\author{\IEEEauthorblockN{
Massinissa Merouani\IEEEauthorrefmark{1},
Afif Boudaoud\IEEEauthorrefmark{1},
Iheb Nassim Aouadj\IEEEauthorrefmark{1},
Nassim Tchoulak\IEEEauthorrefmark{1}\IEEEauthorrefmark{2},
Islem Kara Bernou\IEEEauthorrefmark{1}\IEEEauthorrefmark{2}, \\
Hamza Benyamina\IEEEauthorrefmark{1},
Fatima Benbouzid-Si Tayeb\IEEEauthorrefmark{2},
Karima Benatchba\IEEEauthorrefmark{2},
Hugh Leather\IEEEauthorrefmark{3}, and
Riyadh Baghdadi\IEEEauthorrefmark{1}
}
\IEEEauthorblockA{\IEEEauthorrefmark{1}New York University Abu Dhabi, Abu Dhabi, UAE}
\IEEEauthorblockA{\IEEEauthorrefmark{2}École Nationale Supérieure d'Informatique, Algiers, Algeria}
\IEEEauthorblockA{\IEEEauthorrefmark{3}Meta AI, Menlo Park, CA, USA}
\IEEEauthorblockA{Corresponding authors: massinissa.m@nyu.edu; baghdadi@nyu.edu}
}

\newcommand{\toolName}{\textsc{LOOPer}}
\newcommand\fulldatasetsize{28}
\newcommand\useddatasetsize{10}
\newcommand{\MAPE}{29}
\newcommand{\medianOverPluto}{1.43}
\newcommand{\geomeanOverPluto}{1.42}
\newcommand{\medianOverOldTiramisu}{1.54}
\newcommand{\geomeanOverOldTiramisu}{1.84}

\newcommand{\tablefontsize}{\scriptsize} 

\newcommand{\yes}{\color{listinggreen}{\textbf{\scriptsize Yes}}}
\newcommand{\no}{\color{listingkeywordcolor}{\textbf{\scriptsize No}}}
\newcommand\limited{\color{listingkeywordcolor}{\textbf{\scriptsize Limited}}}

\definecolor{listinggreen}{rgb}{0,0.6,0}
\definecolor{listinggray}{rgb}{0.5,0.5,0.5}
\definecolor{listingmauve}{rgb}{0.58,0,0.82}
\definecolor{listingkeywordcolor}{rgb}{1.0,0.4,0.0}
\definecolor{listinglightgray}{rgb}{0.8863,0.8863,0.8863}

\lstset{ %
  backgroundcolor=\color{white},   
  basicstyle=\linespread{0.7}\footnotesize\ttfamily,        
  columns=fullflexible,
  breakatwhitespace=false,         
  breaklines=true,                 
  captionpos=none,                 
  commentstyle=\color{listinggreen},
  deletekeywords={...},            
  escapeinside={\%*}{*)},          
  extendedchars=true,              
  frame=none,                      
  keepspaces=true,                 
  keywordstyle=\color{listingkeywordcolor}\bfseries,       
  language=C,             
  morekeywords={*,...},            
  numbers=left,                    
  numbersep=5pt,                   
  numberstyle=\tiny\color{listinggray}, 
  rulecolor=\color{black},         
  showspaces=false,                
  showstringspaces=false,          
  showtabs=false,                  
  stepnumber=1,                    
  stringstyle=\color{listingmauve},
  tabsize=2,                       
  title=\lstname                   
}

\maketitle
\thispagestyle{firstpage} 

\begin{abstract}
While polyhedral compilers have shown success in implementing advanced code transformations, they still face challenges in selecting the ones that lead to the most profitable speedups. This has motivated the use of machine learning based cost models to guide the search for polyhedral optimizations. State-of-the-art polyhedral compilers have demonstrated a viable proof-of-concept of such an approach. While promising, this approach still faces significant limitations. Existing polyhedral compilers using deep learning cost models typically support only a small subset of affine transformations, limiting their ability to explore complex code transformations. Furthermore, their applicability does not scale beyond simple programs, thus excluding many program classes from their scope, such as those with non-rectangular iteration domains or multiple loop nests. These limitations significantly impact the generality of such compilers and autoschedulers, raising questions about the overall approach. In this paper, we introduce \toolName{}, the first polyhedral autoscheduler that uses a deep learning based cost model and covers a large space of affine transformations and programs. \toolName{} allows the optimization of an extensive set of programs while being effective at applying complex sequences of polyhedral transformations. We implement and evaluate \toolName{} and show that it achieves competitive speedups over the state-of-the-art. On the PolyBench benchmarks, \toolName{} achieves a geometric mean speedup of \geomeanOverOldTiramisu{}x over the Tiramisu autoscheduler and \geomeanOverPluto{}x over Pluto, two state-of-the-art polyhedral autoschedulers.
\end{abstract}


\section{Introduction}
In an era where compute-intensive applications are everywhere, writing highly optimized code is becoming crucial. Such code is usually manually optimized by experts. This non-trivial task is time-consuming, laborious, and requires profound knowledge of the underlying hardware.

For these reasons, many efforts in the compiler community have focused on developing automatic code optimization frameworks~\cite{adams2019learning,baghdadi2021deep,pencil,bondhugula_practical_2008,polly,Mullapudi:2015:PAO:2786763.2694364,thies_unified_2001,Vasilache2018TensorCF,wolf1991loop,hakimi2023hybrid}.
Polyhedral compilers have shown success in this area~\cite{feautrier_array_1988,Qui00, baghdadi2015PhD,baghdadi2019tiramisu,baghdadi2011speculation, baghdadi2020tiramisuDNNDenseSparse,merouani2020deep}.
They can perform complex iteration space transformations~\cite{wolf1991loop,bondhugula_practical_2008,trifunovic_graphite_2010,polly,Vasilache2018TensorCF,pouchet.11.popl}, data locality optimizations~\cite{Iri88,tobias_hexagonal_cgo13}, and memory management optimizations~\cite{feautrier_array_1988,thies_unified_2001,lefebvre_automatic_1998,Qui00,Darte_contraction_2005}.

Although polyhedral compilers can apply complex program and data layout transformations, they still face challenges selecting the most suitable transformations that yield the best performance~\cite{baghdadi2019tiramisu}. This is primarily due to their search techniques relying on less accurate cost models, leading compilers to make suboptimal decisions.

Previous research has tried to address this problem by using machine learning to build cost models. These cost models are then used to guide the search for code transformations.
Tiramisu~\cite{baghdadi2019tiramisu} is a state-of-the-art polyhedral compiler that has adopted this approach~\cite{baghdadi2021deep}. While it avoids the abovementioned problem, the Tiramisu autoscheduler has significant limitations.

First, the scope of supported programs is limited to programs with single loop nests. In addition, those programs can only have rectangular iteration domains. This critical limitation prevents the Tiramisu autoscheduler from optimizing an important class of programs. In the PolyBench benchmark suite~\cite{louis-noel_PolyBench_2010} for example, 21 out of 30 benchmarks cannot be optimized by the Tiramisu autoscheduler because they are either composed of a sequence of loop nests, or they have non-rectangular iteration domains.
Second, the autoscheduler only explores a small subset of affine transformations (polyhedral optimizations). It only explores five transformations (loop interchange, tiling, parallelization, unrolling, and loop fusion of inner loops within the same loop nest) and each of them can only be applied once. It, therefore, does not cover many important affine transformations. For instance, it does not support loop skewing, which is necessary for parallelizing many stencils. As a result, a stencil, such as a Jacobi 2D with a LARGE\footnote{The LARGE problem size preset as specified by the PolyBench benchmark suite.} input, optimized by the Tiramisu autoscheduler is $10\times$ slower than that optimized by Pluto, which covers a larger space of affine transformations.

In this paper, we propose \toolName{} (a \textbf{L}earned \textbf{O}ptimizer f\textbf{o}r  \textbf{P}olyhedral compil\textbf{er}s), the first deep learning based polyhedral autoscheduler that covers a large space of programs and code optimizations. It explores sequences of affine transformations and uses a deep learning based cost model to evaluate them. \toolName{} supports programs with multiple loop nests as well as programs with rectangular and non-rectangular iteration domains. In addition, it supports a larger set of transformations, including $n$-long sequences of affine transformations. To support these new patterns and enhance performance, \toolName{}'s cost model brings improvements to the cost model and data representation presented~in~\cite{baghdadi2021deep}.
In comparison with other state-of-the-art non-polyhedral autoschedulers that use deep learning (e.g., AutoTVM~\cite{chen2018learning}, Ansor~\cite{zheng2020ansor}, FlexTensor~\cite{zheng2020flextensor}, and Halide\cite{adams2019learning}), \toolName{} is specialized in polyhedral optimizations and therefore can explore complex affine transformations.

To develop \toolName{}, we faced several challenges, including how to effectively explore the space of affine transformations and construct a cost model for this space. Our initial approach involved generating affine schedule matrices and training a cost model to predict performance based on these matrices. However, this was ineffective for two main reasons. First, the space of affine schedule matrices is vast, making exploration time-consuming. Second, developing a precise cost model for all schedule matrices is intractable as it requires an enormous amount of data. We attempted to create such a model, but its accuracy did not meet expectations. This raised crucial questions: how can we efficiently explore this space and improve the model's accuracy while still covering a useful subset of the affine transformation space? The solution is based on the theoretical insight that any unimodular affine transformation can be generated by composing three basic transformations: loop skewing, reversal, and interchange~\cite{wolf1991loop}. This theoretical insight provides two significant advantages: it allows us to cover an interesting subset of affine transformations by composing common transformations, thereby accelerating the search space exploration. Additionally, as we generate regular patterns of affine transformations, the model can learn more effectively with a smaller dataset that we can generate in a reasonable amount of time, enabling the construction of a more accurate model for a subset of the space. We demonstrate that this approach is sufficient to cover an important subset of affine transformations, achieving performance comparable to tools with near-complete modeling of affine transformations, such as Pluto and Pluto+~\cite{bondhugula_practical_2008,10.1145/2896389}.

The contributions of this paper are as follows:
\begin{itemize}
        \item We introduce \toolName{}, the first deep learning based polyhedral autoscheduler that covers a large space of programs and code optimizations. It explores a large set of affine transformations while supporting a large space of programs including programs with multiple loop nests and programs with rectangular and non-rectangular iteration domains.

        \item We release the dataset used to train \toolName{}'s cost model, consisting of \useddatasetsize{} million datapoints. Additionally, we provide the research community with \textsc{LOOPerSet}\cite{looper_data_arxiv25}, an extended dataset of \fulldatasetsize{} million transformed programs, enabling further research in data-driven polyhedral optimization and fostering advancements in machine learning-based autoscheduling.

        \item We implement \toolName{} in the Tiramisu compiler and evaluate it on the PolyBench benchmark suite~\cite{louis-noel_PolyBench_2010}, the gold standard benchmark for polyhedral compilers. We show that \toolName{} achieves a median speedup of $\medianOverOldTiramisu{}\times$ and a geometric mean speedup of $\geomeanOverOldTiramisu{}\times$ over the existing Tiramisu autoscheduler~\cite{baghdadi2021deep}. It also achieves competitive speedups compared to Pluto~\cite{bondhugula_practical_2008} with a median speedup of $\medianOverPluto{}\times$ and a geometric mean speedup of $\geomeanOverPluto{}\times$.
\end{itemize}

\section{Related Work}
This paper proposes an autoscheduler that uses a deep learning cost model to explore affine transformations. In comparison with existing work, \toolName{} has three distinctions:
\begin{itemize}
    \item It supports polyhedral transformations (in contrast to non-polyhedral compilers such as TVM\cite{tvm} and Halide\cite{adams2019learning}).
    \item It uses a deep learning based cost model to evaluate transformations (in contrast to classical polyhedral compilers such as Pluto~\cite{bondhugula_practical_2008}).
    \item It supports a large space of transformations and programs (in contrast to the Tiramisu autoscheduler\cite{baghdadi2021deep})
\end{itemize}

This section first presents autoschedulers used in polyhedral compilers (those that use deep learning and those that do not). Then, we will present non-polyhedral compilers that use deep learning cost models. Table~\ref{tab:related} shows a summarized comparison with state-of-the-art compilers that are the closest to our work (Tiramisu\cite{baghdadi2021deep}, Pluto~\cite{bondhugula_practical_2008}, Halide\cite{adams2019learning} and TVM\cite{chen2018learning}).

\begin{table}[tb]
    \tablefontsize
   
    \setlength\tabcolsep{3pt}
    \caption{Comparison between different autoschedulers.}
    \begin{tabular}{p{3.8cm}|l|l|l|l|l}
        \hline
        
        \textbf{Feature} & \textbf{\toolName{}} & \textbf{Tiramisu} & \textbf{Pluto} & \textbf{Halide} & \textbf{TVM} \\\hline

        \textbf{Deep learning based cost model} & \yes & \yes & \no & \yes & \yes\\\hline
        
        \textbf{Affine transformations} & \yes & \limited & \yes & \no  & \no\\\hline

        \textbf{Rectangular iteration domains} & \yes & \yes & \yes & \yes  & \yes \\\hline

        \textbf{Non-rectangular iteration domains} & \yes & \no & \yes & \limited  & \limited \\\hline

        \textbf{Multiple loop nests} & \yes & \no & \yes & \yes  & \yes \\\hline

        \textbf{Near-complete modeling of affine transformations} & \no & \no & \yes & \no  & \no\\\hline
    \end{tabular}

    \label{tab:related}
\end{table}

\paragraph*{Tiramisu}

The Tiramisu autoscheduler~\cite{baghdadi2021deep}, part of the Tiramisu compiler~\cite{baghdadi2019tiramisu}, uses a tree-based search to explore a fixed set of transformations, including loop interchange, parallelization, tiling, unrolling, and fusion of inner loops within a single loop nest. It relies on an LSTM-based deep learning cost model to guide its search. The model takes features from the unoptimized code and a set of boolean tags representing transformations, recursively embeds the program's Abstract Syntax Tree (AST), and predicts performance from the final embedding.

The current Tiramisu autoscheduler has several limitations. First, its search space is highly restricted: it supports only five transformations, each applied at most once, and notably lacks support for key affine transformations such as skewing, reversal, and shifting.
Second, its reliance on simple boolean tags to represent transformations is not expressive enough to capture ordered sequences of affine transformations. 
Third, it is limited to single loop nests with rectangular iteration domains, restricting its applicability. These limitations mean the Tiramisu autoscheduler supports only one-third of the PolyBench benchmark suite and often fails to apply important optimizations, leading to significant slowdowns. \toolName{} addresses all these issues and adds further enhancements, such as an improved expression representation and support for diverse data types.

\paragraph*{Polyhedral Autoschedulers}

 Polyhedral compilers such as Pluto~\cite{bondhugula_practical_2008}, Pluto+~\cite{10.1145/2896389}, Polly~\cite{polly}, Tensor Comprehensions~\cite{Vasilache2018TensorCF}, PENCIL~\cite{pencil,pencil_paper}, and PolyMage~\cite{Mullapudi:2015:PAO:2786763.2694364} are fully automatic. Some of them are designed for specific domains (such as Tensor Comprehensions and PolyMage), while Pluto, PENCIL, and Polly are more general. Pluto is the most established among them, and its algorithm is used internally by many frameworks~\cite{bondhugula_practical_2008,pencil,polly,Vasilache2018TensorCF}.

\toolName{}'s deep learning cost model is complementary to these compilers, providing a data-driven alternative to their internal heuristics for decisions like parallelism, fusion, and tiling. The primary difference between our approach and Pluto lies in the search methodology. \toolName{} employs a tree search guided by a deep learning cost model to explore its optimization space. Pluto, in contrast, uses an Integer Linear Programming (ILP) solver with a linear objective function to find most affine transformations, complemented by heuristics for remaining decisions.

A key limitation of Pluto's linear objective function is its inability to model complex performance factors like data sizes or target hardware characteristics, which a data-driven model like \toolName{}'s inherently learns from data. Conversely, the simplicity of Pluto's model allows it to cover a near-complete space of affine transformations (especially in Pluto+), a scope that is challenging for a learning-based model to cover accurately given the infinite and complex nature of the space.

Thus, the core tradeoff emerges: \toolName{} prioritizes the precision of a data-driven model on a targeted but powerful subset of transformations, while Pluto prioritizes broader transformation coverage with a simpler, less precise analytical model. We believe that bridging the gap between these two approaches is a promising direction for future research.

\paragraph*{Learning-based Polyhedral Scheduling}
Prior work has also applied machine learning specifically to the polyhedral domain. Park et al.~\cite{5764680} pioneered this by using classic ML models (SVM) to select from a restricted set of high-level primitives. Their approach, however, learned from dynamic features (hardware performance counters) gathered from a baseline execution. In contrast, \toolName{} uses a static program representation and scales the learning problem to a significantly more expressive space of composable affine transformations, guided by a modern deep learning architecture.

More recently, PolyGym~\cite{9563041} explores a complementary direction by formulating the problem as an instance-independent Markov Decision Process for Reinforcement Learning. Its actions abstractly construct a schedule within its geometric state space. This RL formulation differs from \toolName{}’s approach of using supervised learning to train a cost model that guides a combinatorial beam search. Thus, while PolyGym provides the foundational environment for an RL-based solution, \toolName{} is a complete end-to-end system built on the supervised learning paradigm.
\paragraph*{ML-Based Search Methods}

Several state-of-the-art autoschedulers for tensor compilers, such as those in the Halide~\cite{ragan2013halide} and TVM~\cite{tvm} ecosystems, also leverage ML-based search to find efficient schedules.

The Halide autoscheduler~\cite{adams2019learning} pioneered the use of a learned cost model to navigate an image-processing-specific search space. It employs beam search to sequentially construct a schedule, using a feedforward neural network to predict the performance of partially constructed programs and prune the search space. While its search strategy is similar, \toolName{} operates on general polyhedral programs and explores a space of affine transformations not native to the Halide scheduling language.

The TVM stack has a family of increasingly sophisticated optimizers. AutoTVM~\cite{chen2018learning} introduced a template-based auto-tuning approach, using a learned cost model to find optimal parameters within manually-defined schedule templates. To overcome the rigidity of these templates, its successor, Ansor~\cite{zheng2020ansor}, introduced a hierarchical search that generates high-level ``sketches" and then samples low-level parameters to create complete programs. Ansor’s methodology is an online auto-tuning process: it uses a learned cost model to guide an evolutionary search, periodically measures promising candidates on target hardware, and retrains its model. This approach contrasts sharply with \toolName{}, which uses a pre-trained model to make offline decisions without iterative measurement.

A more recent approach, Felix~\cite{felix}, deviates from discrete search algorithms by framing the problem as a differentiable optimization. By defining symbolic schedules with continuous variables, Felix creates a differentiable performance estimator, allowing it to optimize schedules by following the performance gradient. In contrast, \toolName{} operates on a discrete space of polyhedral transformations and relies on a combinatorial search algorithm to explore it.

While these frameworks are powerful for optimizing tensor programs, they do not operate natively within the polyhedral model. As such, their scope is naturally more specialized than \toolName{}'s. They typically lack support for arbitrary affine transformations like skewing, generalized interchange, or reversal, which are essential for many scientific kernels, and they are not designed for programs with non-rectangular domains or in-place memory updates. \toolName{} is explicitly built to handle these cases. Therefore, \toolName{} is a system that targets a fundamentally different and more general class of loop-based programs, making its contributions complementary to advancements in the tensor compiler space.

\section{Overview of the Proposed Approach}

In this paper, we propose \toolName{}, a data-driven polyhedral code optimizer. \toolName{} explores a large space of code transformations and uses a deep learning cost model to guide the search. The design of \toolName{} consists of two main components: the search space exploration module and the evaluation module. These two modules cooperate to optimize the input program.

The role of the search space exploration module is to iteratively build a sequence of code transformations that optimizes a given program. This module consists of two components: a candidate generation algorithm and a search method. The candidate generation algorithm is responsible for suggesting new transformation candidates based on the input program and the current search state. The search method is responsible for defining the space traversal strategy by choosing which candidates to explore next.

The evaluation module is in charge of assessing the quality of candidates that are encountered during the exploration. This module consists of a deep learning model that is trained to predict the potential speedup that sequences of transformations would yield if they were to be applied to the input program.

\subsection*{Scope of This Work}

\paragraph*{Space of supported programs} We are interested in optimizing programs composed of a sequence of rectangular and non-rectangular loop nests that have static affine control. The loop sizes have to be known at compile time. 

\paragraph*{Space of supported transformations} We explore a subspace of the affine transformation space. This subspace is composed of sequences of the following primitive transformations: loop shifting, fusion, skewing, interchange, reversal, parallelization, tiling, and unrolling. Skewing, interchange, and reversal can be applied multiple times and in any order. The maximum length of the transformation sequence is a user-defined hyperparameter.

\paragraph*{Target hardware} We designed our approach to be reproducible on multi-core x86 CPU architectures. In this paper, we implemented and deployed \toolName{} for an Intel Xeon E5-2695 v2 processor.

\section{Search Space Exploration}
\label{exploration}

We structure \toolName{}'s search space as a tree where each node represents a primitive transformation, and a branch is an ordered sequence of transformations. The root of the tree is the untransformed program. Two main components take part in the exploration of the space. First, a candidate generation algorithm that decides, for each node in the tree, what transformations can be applied and appended to the branch. Second, a search method to choose between the candidates and decide which ones to explore further. 


\subsection{Search Method}

First, let us look at \toolName{}'s search method in isolation,
i.e., irrespective of the candidate generation algorithm (which we will present in the next section) and irrespective of the evaluation technique (which we will present in Section \ref{model}). \toolName{} employs beam search to navigate the transformation tree defined by sequences of primitive transformations. Guided by the cost model (Section \ref{model}) acting as the evaluation function, the search maintains and expands only the top \textit{K} candidate transformation sequences (the beam) at each level, where \textit{K} is the beam size. This iterative process terminates when no further valid transformations can be generated from the current beam. Illegal transformations violating data dependencies are detected using standard polyhedral dependence analysis~\cite{feautrier_array_1988,violateddep} and pruned during candidate generation. To avoid local optima, the option of applying no transformation at a given step is always considered as a potential candidate. Cycles are prevented by tracking previously explored transformation sequences.

\subsection{Candidate Generation}

The role of the candidate generation algorithm is to suggest possible expansions of selected branches in the search tree. Given a candidate, the algorithm proposes what transformations to explore next depending on the candidate's position in the search tree, the transformations that have been applied so far, and the characteristics of the input program. The candidate generation algorithm partitions the search space into three levels:

\subsubsection{Fusion Level}

At the first level of the search tree, given the original untransformed program, \toolName{} generates all feasible loop fusion candidates. At this level, we use the loop shifting transformation when needed to enable the fusion of loops that cannot be legally fused otherwise (due to dependencies). 
\subsubsection{Affine Transformation Levels}

After exploring loop fusion candidates and selecting the best ones, we then explore the application of affine transformations. More specifically, we are interested in affine unimodular transformations. 
Concretely, \toolName{} explores $n$-long sequences of interchange, skewing, and reversal, with different parameters, and applied in arbitrary orders. This allows us to cover a large space of affine unimodular transformations while keeping the search space restricted to candidates that are most likely to be profitable.

Given $n$, a parameter that defines the maximum number of affine transformations to be explored,
we create $n$ sub-levels where, at each level, we generate all possible interchange candidates, reversal candidates, and a subset of skewing candidates. We only explore a subset of skewings because the space of possible skewing parameters is infinite. So we sample skewing parameters that can potentially optimize for locality and inner/outer parallelism. We use a Pluto-like algorithm to generate the skewing parameters with the difference being that our algorithm applies to two or three loop levels that will be skewed instead of all the loop levels as in Pluto.

We have explored other methods for generating affine transformations, including a random generation of the affine schedule coefficients. Although this allows the exploration of a wider space of affine transformations, there are two main limitations to this approach:

First, given that the space of affine transformations is infinite, most of the randomly sampled coefficients lead to either illegal or unprofitable transformations \cite{iterative1}. Second, it is harder to build a machine learning based evaluator that can be accurate on all the space of possible affine transformations. This arises from the difficulty of having a large enough training dataset to cover all affine transformations. To avoid these limitations, we restrict \toolName{}'s exploration to sequences of common affine transformations. In practice, we find that this is sufficient.
 
\subsubsection{Final Exploration Levels}

The best candidates from the previous levels are set to be augmented with the following transformations in this order: parallelization, tiling, and unrolling.
We create a sub-level where we generate all possible candidates for each of these transformations.
We generate a candidate for each parallelizable loop, we then explore tiling combinations of perfectly nested loops, and we finish by exploring different unrollings of the innermost loops. Regarding tiling and unrolling parameters, we explore factors from a predefined set of parameters. These are \{32, 64, 128\}  for tiling and \{4, 8, 16\} for unrolling.

\section{Cost Model\label{model}}   

Navigating the large search space requires fast, accurate evaluation of transformation candidates. \toolName{} employs a deep learning cost model, inspired by~\cite{baghdadi2021deep} but significantly extended, to predict the speedup of applying transformation sequences (including shifting, fusion, interchange, skewing, reversal, tiling, parallelization, unrolling) to an input program, replacing costly execution measurements. Our primary contribution is extending the model to handle a much larger space of programs and complex transformation sequences than the original Tiramisu model. In the following sections, we describe \toolName{}'s cost model by briefly explaining the original Tiramisu cost model while highlighting its limitations to better contrast our contributions.

\label{input_representation}
\subsection{Input Representation}

\begin{figure}[t]
            \centering
            \includegraphics[width=\linewidth]{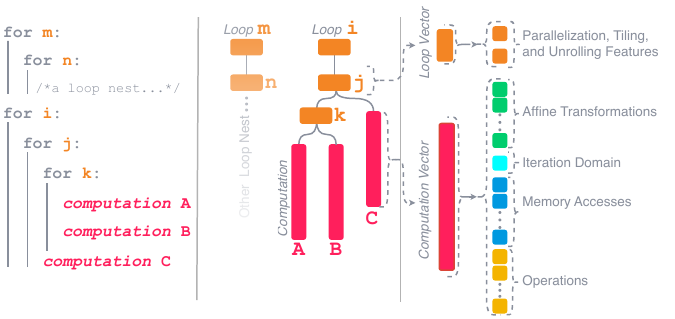}
            
            \caption{Input representation example.}
            \label{fig:input_representation}

\end{figure}

Figure~\ref{fig:input_representation} illustrates \toolName{}'s cost model input representation. We extract high-level features from the program's source and transformation sequence, storing them in a variable-sized format organized as trees mirroring the program's loop nest hierarchy. Leaf nodes represent computations, internal nodes represent loops, and each node holds descriptive features classified as either program or transformation features.

\subsubsection{Program Features}
The program features describe the original untransformed program. Such features include:

    \paragraph*{Computation's expression} We represent expressions via a sequence of vectors encoding operations (add, mul, etc.) from a post-order traversal of the expression tree. This captures structural information crucial for performance modeling, unlike the original Tiramisu model which used only operation histograms, omitting structure. This improved representation enhances prediction accuracy (Section~\ref{ablation_study_summary}).

    \paragraph*{Access matrices} Each statement's memory accesses (reads and writes) are represented by a list of standard polyhedral access matrices. Each matrix captures one access pattern, and includes a numerical identifier specifying which memory buffer is accessed. This allows detailed modeling of data movement.

    \paragraph*{Iteration domain} The original Tiramisu cost model represented iteration domains using loop bounds as integers, sufficient only for its limited scope of rectangular domains. \toolName{} overcomes this limitation by using standard polyhedral representation for iteration domains defining affine constraints over loop iterators. This accurately captures both rectangular and non-rectangular domains directly, proving more effective than approximations like bounding boxes (Section~\ref{ablation_study_summary}).

    To handle these iteration domains in the cost model, we represent each one as an iteration domain matrix, where every row specifies an affine constraint involving the loop iterators and any symbolic constants. We then flatten this integer matrix into a single vector. This vector gets concatenated with the other features that describe the computation, such as its expressions and access patterns (as shown in Figure~\ref{input_representation}). Finally, we pass the combined vector through a fully connected layer (what we call the Computation Embedding Layer in Section~\ref{sec:model_architecture}) to create a unified embedding. This setup allows the model to learn patterns from both rectangular and non-rectangular domains naturally, without any extra steps to treat non-rectangular cases differently.  For example, consider a simple triangular loop nest like 
 \begin{lstlisting}[language=C,escapechar=@]
for (i = 0; i < N; i++)
  for (j = 0; j < i; j++)
    {...}  
\end{lstlisting}
This domain is defined by four inequalities: $i \ge 0$, $N-1-i \ge 0$, $j \ge 0$, and $i-1-j \ge 0$. 
Its iteration domain matrix will look like 
$$\begin{bmatrix} 1 & 0 & 0 \\ -1 & 0 & N-1 \\ 0 & 1 & 0 \\ 1 & -1 & -1 \end{bmatrix}$$  We flatten this matrix into a vector and feed it into the layer alongside other features. This way, the model learns to predict performance based on the exact domain constraints.
    
\subsubsection{Transformation Features}
The second class of features is the transformation features. These features describe the transformations applied. To handle \toolName{}'s expanded search space, particularly sequences of affine transformations (e.g., multiple skewings, interchanges), we represent them as a variable-length list of vectors. Each vector encodes a specific transformation (type and parameters), allowing representation of arbitrary sequences and orderings. This overcomes the limitation of the previous Tiramisu model~\cite{baghdadi2021deep}, which used simple tags suitable only for applying transformations like interchange once. Transformations that can only be applied once per computation remain represented by tags for conciseness. This vector list representation for affine sequences proved equally effective as using full schedule matrices in our ablation study (Section~\ref{ablation_study_summary}) while being conceptually simpler.

\begin{figure*}[t] 
    \centering
    \includegraphics[width=\linewidth]{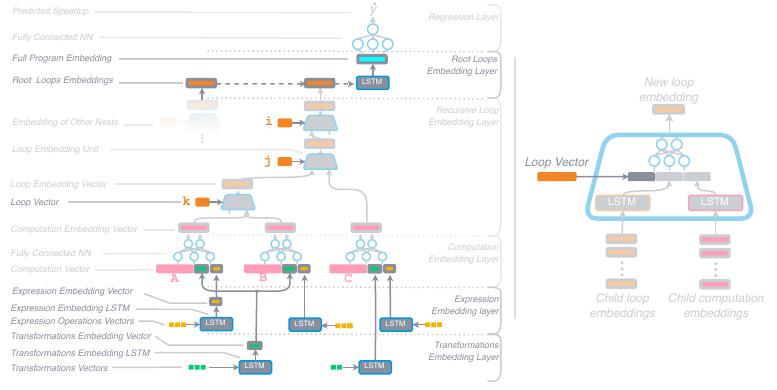}
    
    \subfloat[Processing the tree representation presented in Figure~\ref{fig:input_representation} through \toolName{}'s cost model.\label{fig:model_proc}]{%
        \hspace{0.6\linewidth} 
    }
    \subfloat[Loop embedding unit.\label{fig:loop_unit}]{%
        \hspace{0.3\linewidth} 
    }
    
    \caption{The architecture of \toolName{}'s neural network. The dim-colored elements are parts of the original Tiramisu cost model. The bright-colored parts represent our contributions to the architecture.}
    \label{fig:model_architecture}
\end{figure*}

\subsection{Model Architecture}
\label{sec:model_architecture}

Our cost model formulates speedup prediction as a regression task. Given an input program and a set of transformations, the cost model predicts the expected speedup if these transformations were to be applied. It employs a dynamic AST (Abstract Syntax Tree)-based neural network combining Recurrent and Recursive elements, adapting its structure to each input program's hierarchy (Figure~\ref{fig:model_architecture}). This architecture is inspired by the architecture of the original Tiramisu cost model with some important differences. The original model is composed of three processing layers (as represented by dim-colored parts of the figure). Our proposed model architecture adds three new layers (brightly-colored parts of the figure). These layers allow the model to support our new target space of programs and transformations.
This design emerged as the most accurate and efficient after experimenting with several alternatives (feedforward, LSTM, multi-layer LSTMs, and transformers). Below we describe each layer's function and purpose.

\paragraph*{Affine Transformations Embedding Layer}
The role of this layer is to turn the sequences of affine transformation vectors into a learned embedding. This layer is meant to compress the variable-length sequence of transformations into a fixed-size vector containing an abstract representation of the overall combination of affine transformations. For each computation, this layer takes as input the sequence transformation vectors corresponding to that computation, processes them through an LSTM, and generates an embedding vector for that combination of transformation. This layer did not exist in the original Tiramisu cost model. The addition of this layer enables the support of the application of sequences of transformations of arbitrary length and order.

\paragraph*{Expression Embedding Layer}
The original Tiramisu cost model did not utilize structural information from computation expressions. Instead, it only relied on representing the memory accesses and the operations histogram in a padded fixed-size list.

The purpose of this layer is to generate an embedding that represents the computations' expressions. This embedding is meant to capture information about the type of operations (e.g., add, sub, mul, div, etc.) and their order. This layer takes a representation of the expression operations as input, passes it through an LSTM, and generates an embedding that represents the expressions. 

\paragraph*{Computation Embedding Layer}
This layer is meant to merge the high-level computation features with both the \emph{transformation embeddings} and the \emph{expression embeddings} generated by the previous layers. This layer will then generate a \emph{computation embedding} vector for each computation in the program. This is done by concatenating each computation vector with its corresponding \emph{transformations} and \emph{expression embedding} vectors and then passing the resulting vector through a fully connected neural network.

\paragraph*{Recursive Loop Embedding Layer}
At this level, the entire loop nest and the transformations affecting it get summarized into a single embedding vector. This layer recursively combines \emph{computation embeddings}, loop features, and \emph{loop embeddings} following the hierarchy of the original loop nests. Each recursion step generates a new \emph{loop embedding} that would be fed to the following step. The \emph{root loop embedding} is considered the final embedding of the entire loop nest. At each particular loop level, the embeddings of the child loops (if any) and child computations (if any) are summarized through two different LSTMs, one for the loops and the other for the computations. The resulting vectors, along with the features of the current loop level, are merged through a fully connected neural network, generating a new \emph{loop embedding} vector.

\paragraph*{Root Loops Embedding Layer}
The scope of programs of the original Tiramisu cost model was limited to single loop nests. This restriction prevents the support of a considerable portion of real-world programs. To eliminate this restriction, we introduce a \emph{Root Loops Embedding Layer} that is tasked with aggregating the embeddings of multiple separate loop nests. This layer uses an LSTM to generate the embedding of the entire program given the embeddings of each root loop. 

\paragraph*{Regression Layer}

The final layer of the architecture performs the regression task, predicting the final speedup value. This layer consists of a fully connected neural network that takes as input the final embedding returned by the \emph{Root Loops Embedding Layer} and outputs a single value that represents the predicted speedup.

We selected this AST-based recursive architecture after evaluating alternatives like MLPs, LSTMs, and Transformers, as it proved most effective at modeling the crucial structural and sequential properties of programs for performance prediction. 

\subsection{Data Generation and Model Training}
\label{sec:datagen}

The input space of \toolName{}'s model is very large since it is made of combinations of programs and transformations. Training such a neural network to be accurate in such an ample space requires considerable amounts of labeled data. To produce such a dataset, we generated a corpus of transformed Tiramisu programs for which we measured and recorded the speedup. 

Our data sampling technique involves a two-step process. First, we sample the program space, then we sample the transformation space of each program. The program sampling is done by randomly generating synthetic Tiramisu programs. These programs are generated by combining sequences of basic computation patterns. 

The transformation sampling is done by collecting the candidate transformations encountered by the search technique described in Section \ref{exploration}. We run \toolName{}'s search technique on each synthetic program and, during the exploration, apply each candidate schedule, compile and execute the transformed program, measure its speedup, and store it as a new data point in the dataset. We preferred this sampling approach over random sampling because the latter could include combinations of transformations that are unlikely to be encountered during real exploration. Consequently, this could lead to less relevant examples being incorporated into the dataset. 

Using the proposed data generation technique, we generated thousands of synthetic Tiramisu programs with an average of 133 schedules per program. We experimented with increasing and decreasing the schedules per program ratio and we observed that substantially lower or higher ratios degraded the rate of accuracy improvement per data point generated. This means that, for a fixed data generation budget, increasing the program diversity at the expense of decreasing schedule diversity, or vice-versa, harms the model's accuracy.

The dataset used to train \toolName{}'s cost model comprises 75,000 synthetic programs for a total of \useddatasetsize{} million datapoints. This dataset took approximately ten weeks to generate on a 15-node cluster. Each node in the cluster is equipped with dual 12-core Intel Xeon E5-2695v2 CPUs. In Section \ref{sec:dataset_size_study}, we demonstrate that using just a tenth of this dataset is sufficient for \toolName{} to achieve decent results, slightly outperforming state-of-the-art optimization tools. We also show that training on larger datasets marginally improves performance with diminishing returns. Nonetheless, to address the shortage of performance datasets in the polyhedral compilation research community, we have generated and released \textsc{LOOPerSet}~\cite{looper_data_arxiv25} a larger dataset of \fulldatasetsize{} million datapoints spanning 220,000 synthetic programs. This extended dataset is intended to fuel further research in data-driven polyhedral optimization and to assist projects that are even more data-demanding.

A critical concern when using synthetic data is to ensure the programs are diverse and do not inadvertently replicate the evaluation benchmarks. We therefore conducted a formal diversity analysis on our 220,000 synthetic programs using normalized Tree Edit Distance (nTED). The analysis confirms two key points: first, and most importantly, no program in the PolyBench suite was accidentally replicated. Second, the synthetic programs demonstrate high internal diversity and cover a wide range of structural characteristics, establishing the dataset's quality for training generalizable models. The complete methodology and quantitative results are detailed in our companion paper, ``LOOPerSet: A Large-Scale Dataset for Data-Driven Polyhedral Optimization"~\cite{looper_data_arxiv25}.

Using the \useddatasetsize{} million dataset, we trained \toolName{}'s neural network to predict speedups by minimizing the MAPE (Mean Absolute Percentage Error) loss between the real and the predicted values. Training time on this dataset is 35 hours for 500 epochs. The model was trained on a machine equipped with an AMD EPYC 7742 64-Core Processor and an Nvidia A100 GPU.

\section{Evaluation}

In this section, we demonstrate and analyze the performance of \toolName{} in comparison to state-of-the-art autoschedulers. 
We will first evaluate \toolName{}'s cost model in isolation to assess its reliability as an objective function (Section \ref{sec:model_eval}.) We then evaluate \toolName{}'s efficiency as a complete system in optimizing code (Section \ref{sec:autsched_eval}). This involves comparing the speedups that \toolName{} achieves with those of other autoschedulers, namely Pluto, Pluto+, and the Tiramisu autoscheduler. In the same section, we also evaluate \toolName{}'s search module in isolation by using the ground-truth measurements to guide the exploration. Additionally, we discuss the trade-off between the speed and performance of \toolName{}. In Section~\ref{sec:dataset_size_study}, we will empirically study the influence of the cost model's training set size on the performance of \toolName{}. In Section~\ref{sec:model_portability_study}, we test the portability of the trained cost model on difference CPU microarchitectures. Finally, we conclude the evaluation by providing an ablation study on the model's architecture to justify our design choices (Section~\ref{ablation_study_summary}).

\paragraph*{Experimental Setup}
We performed the evaluation on an Ubuntu 22.04.3 system running on a dual-socket 12-core Intel Xeon E5-2695v2 CPU equipped with 128 GB of RAM. For all experiments, we used \texttt{GCC} (version 11.4.0) as the backend C/C++ compiler. While Intel compilers are a common choice for evaluating performance on Intel CPUs, our decision was based on preliminary empirical results. We compared the performance of Pluto-optimized PolyBench codes compiled with \texttt{GCC} versus the Intel oneAPI DPC++/C++ Compiler (\texttt{icpx} 2023.2.0). Overall, we found no consistent advantage for either compiler; in fact, the geometric mean performance of code compiled with \texttt{icpx} was $0.85\times$ that of \texttt{GCC}. While \texttt{icpx} led on 36\% of benchmarks, \texttt{GCC} was superior on 45\%, with the remainder being comparable. Given these mixed results and our goal of maintaining a simple and widely accessible experimental setup, we proceeded with \texttt{GCC} for all reported evaluations.

\begin{figure}[t]
    \centering
    \includegraphics[width=\linewidth]{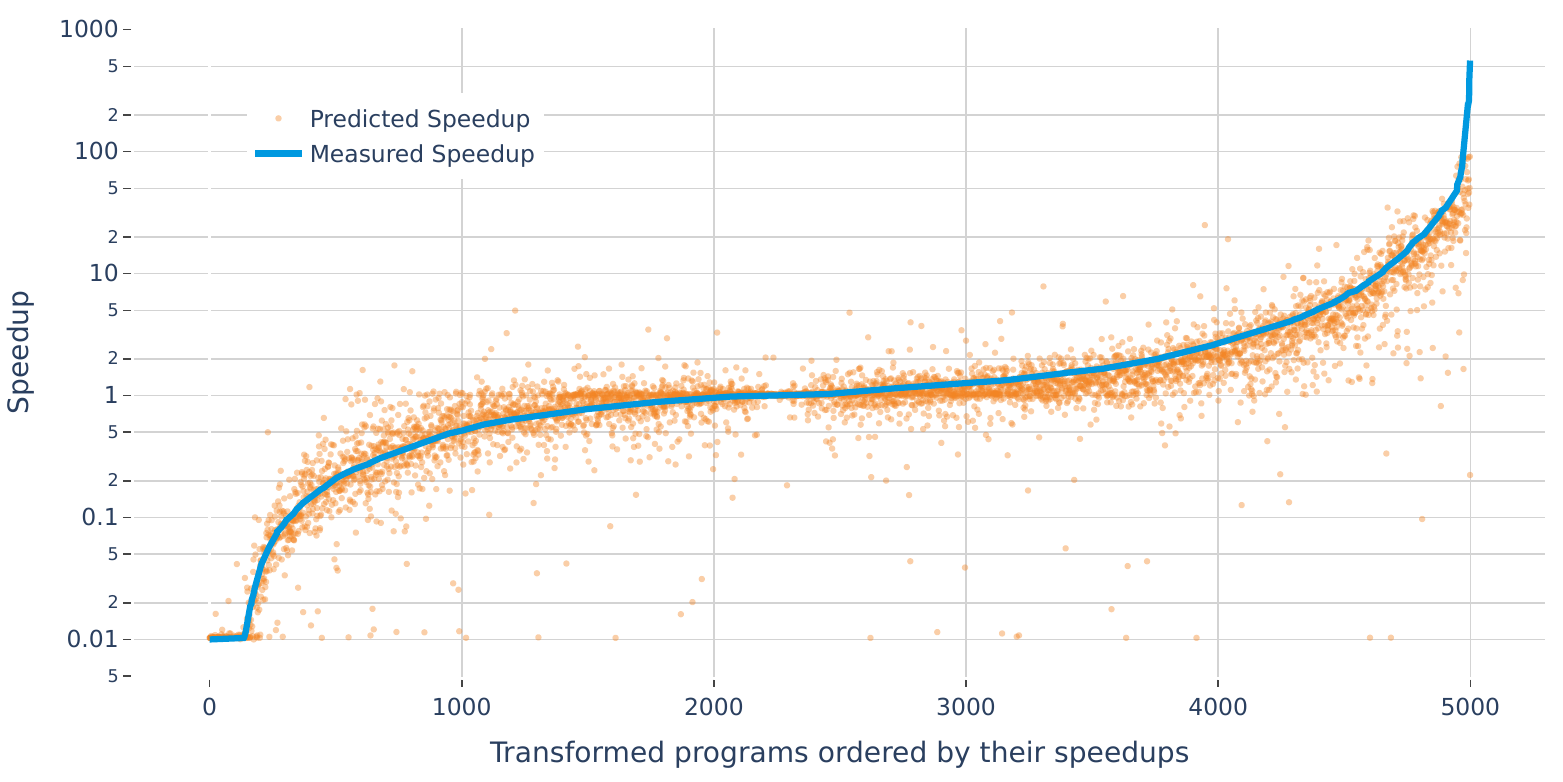}
\caption{Predicted speedups compared to measured speedups}
\label{fig:predicted_vs_real}
\end{figure}

\subsection{Cost Model Evaluation}
\label{sec:model_eval}

We evaluate the prediction accuracy of \toolName{}'s cost model by computing relevant metrics on a test set of unseen programs. To train and evaluate our model, we split our dataset into a training set (90\%) and a validation set (10\%). To ensure the soundness of our testing, we make sure that no program is repeated in both the validation and the training set. Key metrics demonstrate its effectiveness: the Mean Absolute Percentage Error (MAPE) is $\MAPE{}\%$; the Spearman’s rank correlation coefficient between predicted and measured speedups is 0.75, indicating strong rank correlation; and the average Normalized Discounted Cumulative Gain (nDCG) score is 0.96, signifying near-ideal ranking ability for identifying promising transformations. While the MAPE is higher than the previous Tiramisu model's~\cite{baghdadi2021deep}, this reflects the significantly more complex program and transformation space \toolName{} handles, yet, as shown later, still enables substantial speedups. 
Figure \ref{fig:predicted_vs_real} compares predicted and measured speedups. We use a random subset of the test set consisting of 5000 transformed programs. To simplify the visualization, we sort the transformed programs based on their speedups in ascending order. The figure shows that the predicted speedups are close to the measured ones.

\subsection{Evaluation of the Autoscheduler}
\label{sec:autsched_eval}

\paragraph*{Evaluation Benchmarks}

 To evaluate \toolName{}'s autoscheduling performance, we use the PolyBench benchmark suite~\cite{louis-noel_PolyBench_2010}, the gold standard benchmark suite for polyhedral compilers. PolyBench consists of 30 benchmarks that are extracted from various computing areas, including linear algebra, stencils, physics simulation, etc. We used version 4.2.1\footnote{https://polybench.sf.net}. For each benchmark, we used the five problem sizes that PolyBench defines (MINI, SMALL, MEDIUM, LARGE, and EXTRALARGE) and the default PolyBench data types. To simplify the presentation of the results, we take the geometric mean of the speedups obtained on all five sizes for each benchmark. It is important to note that benchmark programs were not used to train the cost model. \toolName{}'s cost model is exclusively trained offline on randomly synthesized programs, as explained in Section~\ref{sec:datagen}.

\begin{figure*}[t]
\centering
\includegraphics[width=\linewidth]{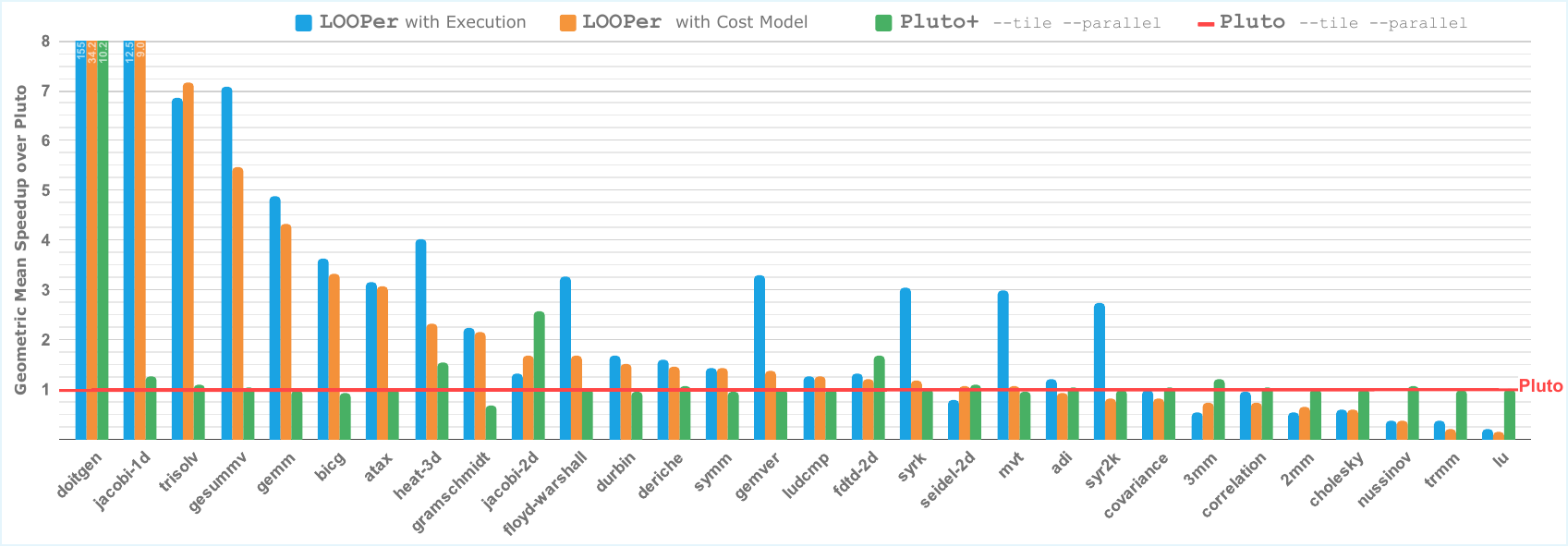}
\caption{Speedups of \toolName{} (using the cost model and using the actual measurements) compared to Pluto and Pluto+. The speedups are aggregated by geometric mean over the five sizes of each benchmark. The benchmarks are sorted by descending order of \toolName{}'s speedups.}
\label{fig:plutoVsUs}
\end{figure*}

\subsubsection{\textbf{Comparison with Pluto\label{plutoexp}}}

This section compares \toolName{} with the Pluto autoscheduler~\cite{bondhugula_practical_2008}. Figure \ref{fig:plutoVsUs} shows the speedups for both on the PolyBench benchmarks. We used Pluto with the options {\small \verb|--parallel --tile|} to enable parallelism and tiling in addition to default transformations (which include affine transformations and fusion). 
The baseline for calculating speedups is the execution time obtained using Pluto. The column \emph{``LOOPer with Cost Model"} represents the speedups found by \toolName{} using the cost model to guide the exploration (i.e., our proposed approach). These results are achieved using a beam of size $3$ in \toolName{}'s search method.

\toolName{} matches or outperforms Pluto in 20 benchmarks out of 30, achieving a median speedup of \medianOverPluto$\times$ and a geometric mean of \geomeanOverPluto$\times$. This is mainly because \toolName{}  uses a more precise cost model that takes into consideration multiple factors (all of which are data-driven). Pluto, in contrast, uses a linear objective function that tries to minimize the distance between producer and consumer statements to improve data locality and enable parallelism. However, Pluto's objective function does not consider many important performance-related factors.

One important factor that our model takes into account, but Pluto ignores, is the computational and memory workload of each loop. This workload depends on many factors such as the loop extents, nesting depth, nature and number of computational operations, number of memory accesses, and the spatial and temporal distance between memory accesses, etc. Considering this aspect is crucial when deciding whether to parallelize a program and, if so, which loop to parallelize. Our tool, \toolName{}, only parallelizes a loop if the benefit of parallelization outweighs the overhead it introduces. This decision is entirely data-driven, as we did not need to engineer \toolName{} to handle this specific case. In contrast, Pluto always parallelizes the outermost loop when it is legal, which is not always the optimal strategy and can often have adverse effects in many scenarios.

A clear example of this strategy's limitations is seen in the \emph{jacobi-1d} benchmark. Pluto applied parallelization after fusing the two computations and skewing the loop nest, while our approach opted to skip parallelization across all problem sizes. Even for larger sizes, \toolName{} accurately determined that parallelization would degrade performance and instead restrained to applying fusion, skewing, and unrolling. This demonstrates that, despite the sizable iteration domain and loop extents, \toolName{} accounted for other factors influencing transformation profitability, such as computational load. In the case of \emph{jacobi-1d}, the shallowness of the loop nest and the simplicity of the computations justified the decision to avoid parallelization, resulting in schedules that achieved an average (in the geometric mean sense) speedup of $9\times$ over Pluto.

Another advantage of \toolName{} is its ability to tailor different schedules for the different sizes of the same program. In contrast, Pluto is size-agnostic, generating the same schedule for a benchmark regardless of loop extents or iteration domain size. The same schedule cannot necessarily fit all the sizes of a program. For instance, on the \emph{heat-3d} benchmark, Pluto decided to apply a combination of fusion, skewing, parallelization, and 4D tiling to all sizes. However, this schedule happens to deteriorate the performance of (i.e. slows down) 4 out of the 5 sizes compared to the {\small \verb|gcc -O3|} version. \toolName{} proposed a unique schedule for each size, using combinations of 2D tiling, parallelization, interchange, and unrolling, leading to consistent outperformance or parity with {\small \verb|gcc -O3|} across all sizes. This translates to $2.3\times$ speedup of \toolName{} compared to Pluto on \emph{heat-3d}. 

The importance of a comprehensive cost model is further highlighted in the \emph{symm} benchmark. Here, \toolName{}'s cost model determined that opting out of applying any transformation is better than any of the explored schedules, and this indeed results in a $1.4\times$ speedup compared to Pluto's heuristic.

Pluto outperforms \toolName{}  in 10 out of 30 benchmarks. This is typically due to model mispredictions that misguides the search or to the fact that Pluto leverages transformations that we do not explore.
In the \emph{syrk\_LARGE} benchmark, for example, our model wrongly predicted that a tiling would improve the parallelization of the program whereas an interchange would have done a better job. Although this is a good schedule (a speed up of $109\times$ compared to the untransformed version), Pluto outperformed \toolName{} and produced a transformation that is  $1.2\times$ faster than ours.
Another advantage to Pluto is that it applies some transformations that were not reached by \toolName{}. For example, in both \emph{nussinov} and \emph{cholesky}, \toolName{} could not apply any transformations, whereas Pluto was able to modify the code through affine transformations, enable parallelization, and outperform our system. In those two benchmarks, Pluto was able to segregate non-parallelizable computations into separate loop nests using loop distribution. These nests are run sequentially while the rest of the computations are parallelized. Enabling parallelization in these two specific cases would require transformations like loop distribution (also known as loop fission), which are not currently supported by \toolName{} but are set to be added in future work.


\subsubsection{\textbf{Comparison with Pluto+}}
An improved version of Pluto exists, Pluto+~\cite{10.1145/2896389}. The latter is a slightly stronger baseline; however, we decided to frame the primary comparison against Pluto given its status as a widely recognized and foundational tool in the polyhedral community, which helps contextualize our results. 
We also compare \toolName{} with Pluto+ as shown with a green bar in Figure~\ref{fig:plutoVsUs}. In our test setup, the results for Pluto+ vs. \toolName{} are close to the ones presented in the previous section (Sec. \ref{plutoexp}), with the geometric mean on all of PolyBench being $1.27\times$ and a median of $1.36\times$. Against this baseline, \toolName{} matches or outperforms Pluto+ in 18 benchmarks out of 30. In our experiments, Pluto+ gives a $15\%$ performance improvement over Pluto in PolyBench, which is not enough to bridge the gap between \toolName{} and Pluto+.

\subsubsection{\textbf{Comparison with Measurement-guided Exploration}}
\label{execVsModel}

In this section, we compare the performance of \toolName{} using two different evaluation methods: first, using the cost model as an evaluation function, and second, by compiling and executing those candidates to get their execution time (i.e. ground-truth speedup measurements).
Figure \ref{fig:plutoVsUs} shows the speedups obtained by the two methods using a beam of size 3.
The column \emph{``LOOPer with Execution"} shows the results of using \toolName{} with execution instead of using the cost model. Results obtained by execution (a perfect model) represent the maximum speedups \toolName{} could achieve with the current search method. The better our cost model is, the closer we are to these ideal speedups. 

The downside of guiding search with execution is that it requires compiling and executing every candidate transformation encountered during the search. This significantly slows down the search and can render it impractical for large programs or large search spaces. Using a cost model provides a compromise between the search time and the quality of the schedules found.

In many benchmarks, the autoscheduler guided by the cost model is able to achieve comparable results to the autoscheduler guided by execution, with an overall median ratio of $0.92$ and a geomean of $0.75$ between the two (speedup by model/speedup by execution). This disparity is justified by the fact that the cost model's predictions are not perfect and this can mislead the search into lesser quality solutions (as explained in Section~\ref{Search Speed of Model vs Execution}).
This can be seen in the \emph{gemver} benchmark, where the speedup obtained by \toolName{} represents only $41\%$ of the speedups obtained by execution.

In some cases like \emph{seidel2d\_LARGE}, \toolName{} achieves a higher speedup when guided with the cost model than when measurements guide it. The greedy nature of beam search can prevent the measurements-guided exploration from reaching some profitable transformations that the model-guided exploration was able to reach. This happens when the cost model misranks schedules that have close ground-truth speedups leading it to explore paths not explored by the measurements-guided exploration.

\subsubsection{\textbf{The Search Speed Trade-off}}
\label{Search Speed of Model vs Execution}

 Using a cost model to evaluate different optimization candidates instead of ground-truth measurements is useful for two reasons: first, this allows faster space exploration. This is because predicting speedups using the cost model is significantly faster than compiling and running programs (especially for programs that have large inputs). The second reason is that in some contexts, cross-compilation is necessary, and compiling on the target machine is not easy; therefore, allowing the compiler to optimize code even in the absence of access to the target machine is desirable.

 This section compares the search time achieved with a model-guided exploration and a measurement-guided one (i.e., with actual measurements after compiling and executing transformed programs). To do so, we run \toolName{} with both evaluation methods and with the same beam size (a beam size of 3) on the entire PolyBench set, and we record the search times for each program.
 We find that the model-guided \toolName{} is, on average, $644\times$ faster than the measurement-guided \toolName{} while delivering comparable results as shown previously (Section \ref{execVsModel}). This speed difference is justifiable by the fact that the inference time of our model is $32$ms on average (inference on CPU for non-batched input) whereas compiling and executing a candidate can be orders of magnitude slower.
 Taking the benchmark \emph{seidel2d\_LARGE} as an example, our method's search time is $25.61$ seconds. Using the measurement-guided exploration, the search time is more than $9$ hours, which means that the search speed has improved by a factor of $1265\times$. 

 \begin{figure}[t]
    \centering
    \includegraphics[width=\linewidth]{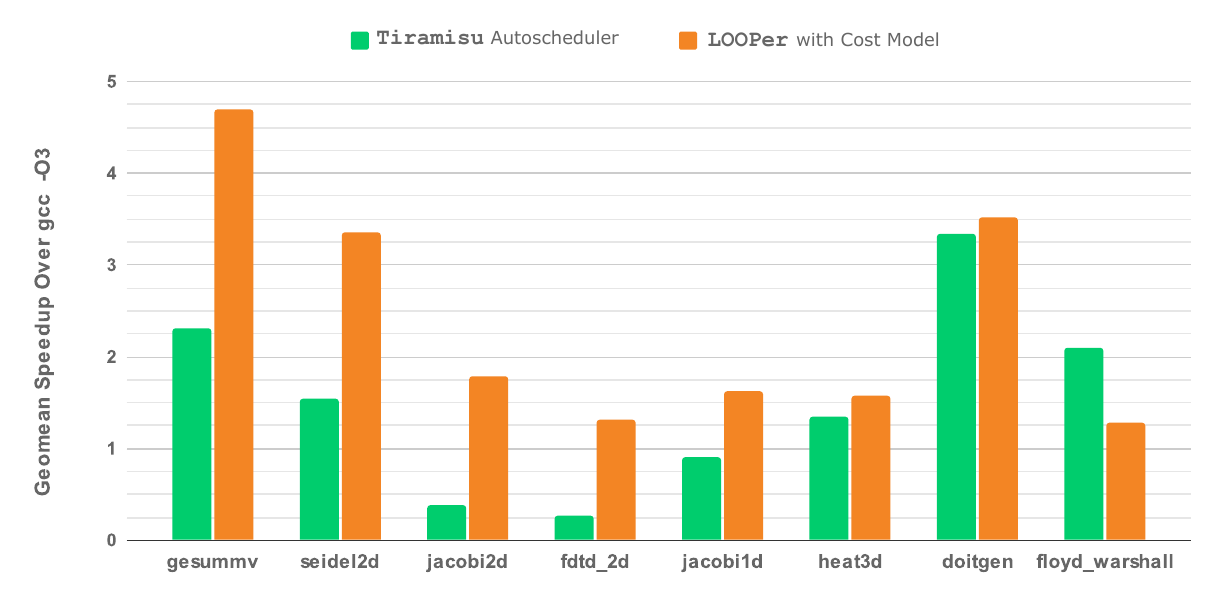}
    \caption{Speedups of \toolName{} compared to the Tiramisu autoscheduler. The speedups are aggregated by geometric mean over the five sizes of each benchmark that the Tiramisu autoscheduler supports.}
    \label{fig:OldVsHalideVsUs}
\end{figure}

\subsubsection{\textbf{Comparison with the Tiramisu autoscheduler}}

In this section, we compare \toolName{} with the Tiramisu autoscheduler described in \cite{baghdadi2021deep}. We use eight benchmarks from PolyBench for this comparison because the Tiramisu autoscheduler does not support the rest of PolyBench due to two main limitations: not supporting non-rectangular iteration domains and not supporting programs with multiple loop nests. For the case of the \emph{gemm} benchmark, it is supposed to be supported by Tiramisu, but the autoscheduler crashes when we try this particular benchmark. 

With a geometric mean speedup of \geomeanOverOldTiramisu$\times$ and a median speedup of \medianOverOldTiramisu$\times$, \toolName{} outperforms Tiramisu in 7 out of 8 benchmarks, as shown in figure \ref{fig:OldVsHalideVsUs}. The difference in speedups is because \toolName{} considers a larger space of affine loop transformations. It supports the application of shifting, loop fusion at any loop level, and the application of multiple skewings, reversals, and interchanges. In contrast, the Tiramisu autoscheduler does not support shifting, skewing, and reversal and can apply interchange only once. It also does not support the application of loop fusion in its general form.

In the benchmarks, loop skewing is the most important affine transformation exploited by \toolName{} and not exploited by Tiramisu's autoscheduler.
Skewing not only improves data locality but it can also enable parallelism. A clear example is the \emph{jacobi2d} benchmark, where parallelization is not legal without skewing. In this specific benchmark, \toolName{} outperforms the Tiramisu autoscheduler on all sizes and aggregates to being $4.57\times$ faster.  

\toolName{} inaccurately predicted some transformations as beneficial in cases where simpler ones would have been enough. This has allowed Tiramisu to get better speedups in \emph{floyd\_warshall}. This problem arises because the cost model of the Tiramisu autoscheduler is more accurate. It has a MAPE of $16\%$ compared to $\MAPE\%$ for our model. This difference can be explained by the fact that \toolName{}'s model covers a significantly larger space of transformations and programs, making speedup prediction much more difficult.

\subsection{Influence of the Training Set Size on the Cost Model's Performance}
\label{sec:dataset_size_study}
\begin{figure}[t]
        \centering
        \includegraphics[width=\linewidth]{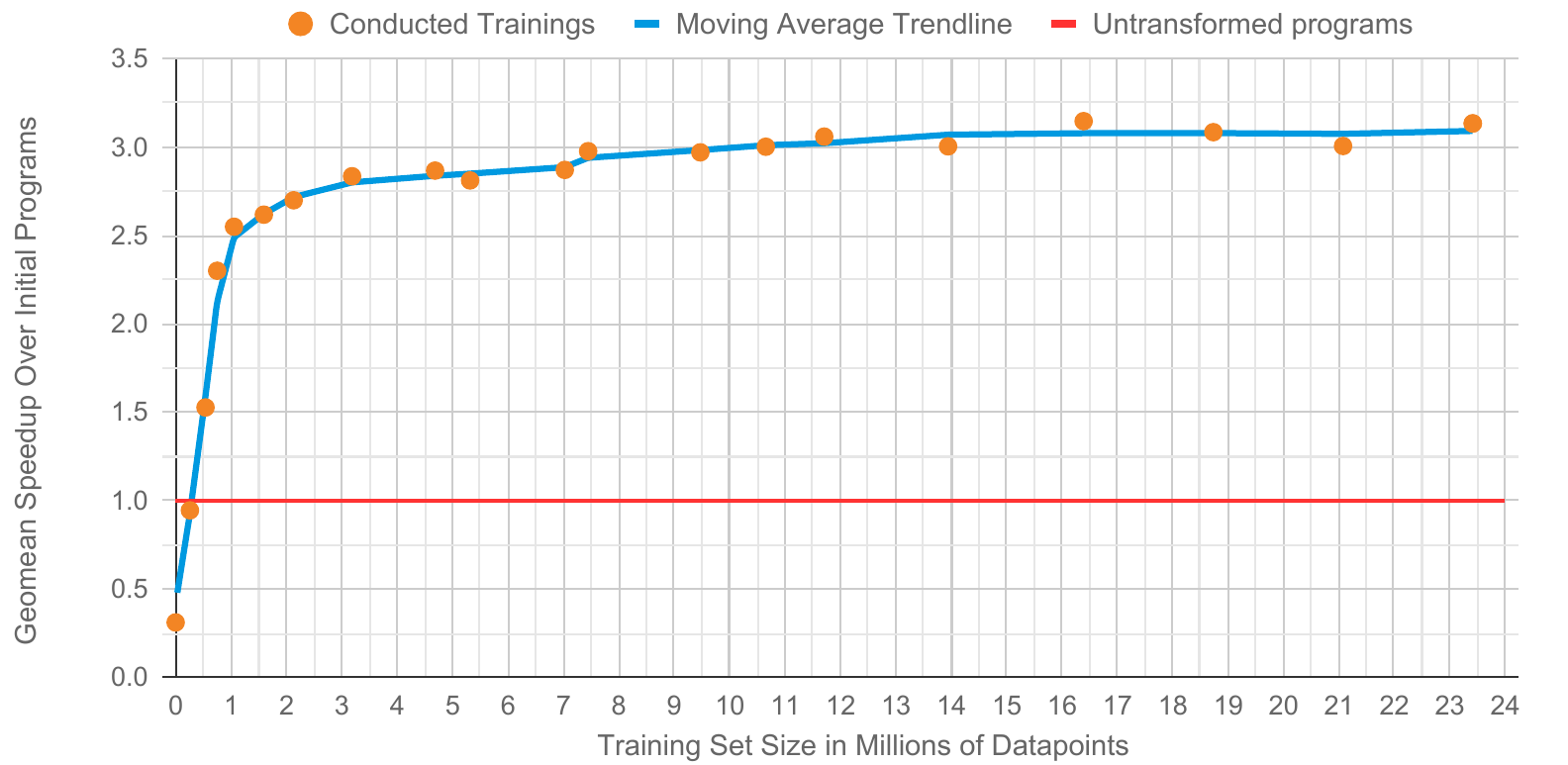}
        \caption{ Evolution of the geometric mean speedup over the training set size.}
        \label{fig:sp_over_ds_size}


\end{figure}

A key question in designing a data-driven approach for code optimization is determining the right amount of training data needed to achieve effective results. The answer can depend on several factors, including the dimensionality of the program/transformation space, the data generation technique, and the architecture of the model.

For \toolName{}, we sought to estimate the minimum amount of data required for decent performance by training multiple instances of its cost model on datasets of incremental sizes, ranging from 250 thousand datapoints up to 23.5 million. For each subset, we trained a separate model instance from scratch and used it for autoscheduling. During this process, we recorded both the autoscheduling performance and the model's error rate.
To assess autoscheduling performance, we evaluated the speedup achieved by \toolName{} on 1,000 randomly selected synthetic programs that were excluded from the training sets. In this experiment, the speedup is defined as the ratio of the program's original execution time to its execution time after \toolName{}'s optimization. The results are aggregated using the geometric mean. We also reported the Mean Absolute Percentage Error (MAPE) of each cost model instance on a fixed validation set that remained consistent across all experiments.

Figure \ref{fig:sp_over_ds_size} illustrates the evolution of the geometric mean speedup as the training set size increases. With an untrained model (training set size of 0), \toolName{} achieves a geometric mean speedup of $0.31\times$ (or a $3.2\times$ slowdown), as the search behaves similarly to random exploration. As the training set grows, performance improves with diminishing returns, eventually converging around a speedup of $3\times$. Similarly, 
we find 
a corresponding negative exponential trend in the MAPE on the validation set, with diminishing returns as the training set size increases. Initially, the MAPE is close to $100\%$ for the untrained model, which occurs because the model tends to predict values near zero due to our weight initialization approach. As the training progresses, the MAPE curve stabilizes around $24\%$.

Based on the results in Figure \ref{fig:sp_over_ds_size}, we selected a training set of 9 million data points, as the performance gains beyond this size were marginal. Cost model  used throughout this paper was developed using \useddatasetsize{} million datapoints (9M for training and 1M for validation), resulting in a geometric mean speedup of $\geomeanOverPluto{}\times$ over Pluto on PolyBench.


Further experiments indicate that training the model with as few as 1 million datapoints is sufficient to marginally outperform Pluto, achieving a geometric mean speedup of $1.1\times$ on PolyBench.

To support further research, we made the entire dataset generated for this project (comprising \fulldatasetsize{} million datapoints) publicly available to the compiler research community~\cite{looper_data_arxiv25}.


\subsection{Cost Model's Portability to Different CPU Microarchitectures}
Apart from the cost model being trained specifically for the CPU on which the dataset was generated, the rest of the method is microarchitecture-agnostic. \toolName{}’s search space design, program characterization, and cost model architecture do not rely on any CPU-specific features. Given that the cost model was trained using performance data from a specific CPU, an important question arises: can this model be reused, without retraining, for autoscheduling on other CPUs? To investigate this, we evaluated the cost model, without retraining, on a variety of CPUs with differing levels of similarity to the original training CPU. Table~\ref{tab:cpus_specs} outlines the specifications of the CPUs employed in our portability experiments.

\begin{table}[t]
\tablefontsize
\caption{\tablefontsize List of CPUs used for portability experimentation}
\label{tab:cpus_specs}
\begin{tabular}{@{}lcccc@{}}
\toprule
                  & \begin{tabular}[c]{@{}c@{}}\emph{CPU1}\\ (training CPU)\end{tabular}    & \emph{CPU2}                                                             & \emph{CPU3}                                                     & \emph{CPU4}                                                     \\ \midrule
Model             & \begin{tabular}[c]{@{}c@{}}Intel Xeon \\ E5-2695 v2\end{tabular} & \begin{tabular}[c]{@{}c@{}}Intel Xeon \\ E5-2680 v3\end{tabular} & \begin{tabular}[c]{@{}c@{}}AMD EPYC \\ 7R13\end{tabular} & \begin{tabular}[c]{@{}c@{}}AMD EPYC \\ 7742\end{tabular} \\ \midrule
Microarchitecture & Ivy Bridge-EP                                                    & Haswell                                                          & Zen 3                                                    & Zen 2                                                    \\ \midrule
Family            & Xeon E5                                                          & Xeon E5                                                          & EPYC 7002                                                & EPYC 7003                                                \\ \midrule
Sockets count     & 2                                                                & 2                                                                & 2                                                        & 2                                                        \\ \midrule
Cores per socket  & 12                                                               & 12                                                               & 48                                                       & 64                                                       \\ \midrule
Threads per core  & 2                                                                & 2                                                                & 2                                                        & 1                                                        \\ \midrule
Cache (L3)        & 60 MB                                                            & 60 MB                                                            & 384 MB                                                   & 256 MB                                                   \\ \bottomrule
\end{tabular}
\end{table}

To quantify changes in autoscheduling performance, we performed both measurement-guided autoscheduling (\toolName{} with execution) and model-guided autoscheduling (\toolName{} with cost model) on each of the CPUs and compared the results. In this experiment, the measurement-guided autosheduling has the ability to take different search paths for each CPU since it evaluates the candidates on the target CPU itself. In contrast, model-guided autoscheduling lacked performance insights from the new CPUs since it is trained solely on data collected from \emph{CPU1}. 

Table~\ref{tab:cpu_speedup_comparison} summarizes performance ratios (execution time of the best schedule found by measurement-guided \toolName{} divided by that of the model-guided \toolName{}) for the entire PolyBench suite.

\begin{table}[t]
\tablefontsize
\caption{\tablefontsize Speedup of best schedules found by \toolName{} with (non-retrained) model over \toolName{} with execution on PolyBench}
\label{tab:cpu_speedup_comparison}
\centering
\begin{tabular}{@{}lllll@{}}
\toprule
               & \emph{CPU1}         & \emph{CPU2}         & \emph{CPU3}         & \emph{CPU4}         \\ \midrule
Geometric mean & $0.75\times$ & $0.74\times$ & $0.72\times$ & $0.68\times$ \\ \midrule
Median         & $0.92\times$ & $0.93\times$ & $0.91\times$ & $0.83\times$ \\ \bottomrule
\end{tabular}
\end{table}

We observed a slight performance decline on \emph{CPU2}, decreasing from $0.75\times$ to $0.74\times$. This minimal difference is attributed to the similarity between the model's training CPU and \emph{CPU2}. These two CPUs are from the same family but one generation apart, they therefore have comparable performance response to transformations. The performance degradation on \emph{CPU3} is about $3\%$, which is reasonable given that these CPUs are from different manufacturers and have different microarchitectures. A notable decrease of about $7\%$ was observed with \emph{CPU4}, possibly attributable to the simultaneous multithreading (SMT) being disabled in this case, which significantly affects the performance behavior of transformations on this CPU.

This experiment shows that the cost model, even without fine-tuning, performs reasonably well on similar microarchitectures. However, we hypothesize that retraining or fine-tuning the cost model on the new CPUs would improve performance further.

\label{sec:model_portability_study}

\subsection{Cost Model Ablation Study Summary}
\label{ablation_study_summary}
Ablation studies confirmed the benefits of our model's input representations. Using precise polyhedral iteration domains improved accuracy over rectangular approximations (4.6\% lower MAPE). Our structured expression representation modestly outperformed simple operation histograms (1\% lower MAPE). Representing transformation sequences as vectors was as effective as schedule matrices while being more concise. 

\section{Limitations and Future Work}

While \toolName{} has shown promising results compared to state-of-the-art, it still has limitations. Currently, the search space we cover does not include loop distribution. In addition, while certain affine transformations (skewing, interchange, and reversal) are explored in an arbitrary order, the other transformations are explored in a fixed order.
While \toolName{}'s cost model, as presented in this paper, is fit specifically to the CPU on which data was collected, the approach itself is hardware-independent and can be reproduced for other CPU architectures without any adaptations required. This ease of portability is because all parts of the exploration technique, the model’s architecture, and the input characterization are hardware-independent. In order to port \toolName{} to significantly different CPU architectures, one only needs to generate a new dataset for the target CPU and retrain or fine-tune the cost model. 
Supporting different CPUs is left for future work.
Future research directions will focus on addressing these limitations and improving the model's accuracy.

\section{Conclusion}

In this paper, we presented \toolName{}: a novel polyhedral autoscheduler that explores affine transformations using a deep learning based cost model. Compared to state-of-the-art, this autoscheduler covers a large subset of affine transformations, allowing the application of complex sequences of polyhedral transformations. It also supports programs that have multiple loop nests as well as programs that have rectangular and non-rectangular iteration domains, allowing the optimization of a larger set of programs.

\toolName{} shows that it is possible to use deep learning to predict speedups for a large space of affine transformations and programs, opening the door for further  research into the use deep learning based cost models.

The proposed cost model has a MAPE of \MAPE\% and achieves a median speedup of \medianOverPluto$\times$ over Pluto, a state-of-the-art polyhedral compiler. It achieves a median speedup of \medianOverOldTiramisu$\times$ over the Tiramisu autoscheduler, mainly because it can explore a larger space of affine transformations.

\section*{Acknowledgment}
This research has been partly supported by the Center for Artificial Intelligence and Robotics (CAIR) at New York University Abu Dhabi, funded by Tamkeen under the NYUAD Research Institute Award CG010. A part of the research was carried out on the High-Performance Computing resources at New York University Abu Dhabi.
The authors are grateful for the considerable computational resources provided by the Commit research group, led by Professor Saman Amarasinghe at the MIT Computer Science and Artificial Intelligence Laboratory (CSAIL).

\newpage

\bibliographystyle{IEEEtran}
\bibliography{sample-base}

\appendices

\section*{Artifact Evaluation Appendix}

\subsection{Abstract}

This artifact accompanies the paper ``LOOPer: A Learned Automatic Code Optimizer For Polyhedral Compilers". It contains the source code for the LOOPer autoscheduler, which is integrated into the Tiramisu compiler, along with the pre-trained deep learning cost model, the PolyBench benchmark suite, and all necessary evaluation scripts. The entire environment is packaged in a Docker container to ensure ease of use and full reproducibility of the software environment. The artifact enables the reproduction of the key performance results from the paper, specifically the speedup comparison between LOOPer and the Pluto compiler (Figure 4), by running a set of automated scripts.

\subsection{Artifact check-list (meta-information)}

{\small
\begin{itemize}
  \item {\bf Algorithm: } Learned polyhedral autoscheduling, Beam search with a deep learning cost model.
  \item {\bf Program: } LOOPer autoscheduler, Tiramisu compiler, PolyBench v4.2.1 benchmark suite.
  \item {\bf Compilation: } g++, cmake. All dependencies are handled by the provided Docker container.
  \item {\bf Transformations: } Polyhedral transformations: skewing, interchange, reversal, fusion, tiling, parallelization, unrolling.
  \item {\bf Binary: } The LOOPer toolchain is pre-compiled in the base Docker image.
  \item {\bf Model: } Deep learning cost model (LSTM-based recursive neural network). Pre-trained weights are provided.
  \item {\bf Dataset: } PolyBench v4.2.1 (for evaluation). A 10M-point training set as part of the LOOPerSet dataset \url{https://huggingface.co/datasets/Mascinissa/LOOPerSet}.
  \item {\bf Run-time environment: } Docker container (Ubuntu 22.04, Python 3.10, PyTorch). No root access required.
  \item {\bf Hardware: } Multi-core x86-64 CPU. Paper's results are from an Intel Xeon E5-2695 v2; performance will vary on other hardware.
  \item {\bf Run-time state: } Performance measurements are sensitive to concurrent system load; experiments should be run on an idle machine.
  \item {\bf Execution: } Automated via shell scripts (\texttt{run\_comparison.sh}).
  \item {\bf Metrics: } Execution time, Speedup (geometric mean).
  \item {\bf Output: } CSV files, summary tables printed to console.
  \item {\bf Experiments: } Reproduce performance comparison of LOOPer vs. Pluto on PolyBench (Figure 4).
  \item {\bf How much disk space required (approximately)?: } 10 GB (for the Docker image).
  \item {\bf How much time is needed to prepare workflow (approximately)?: } $<$ 15 minutes (to build the Docker container).
  \item {\bf How much time is needed to complete experiments (approximately)?: } Kick-the-tires: $\sim$30 min. Main results: $\sim$3-4 hours. Full results: \textgreater24 hours.
  \item {\bf Publicly available?: } Yes.
  \item {\bf Code licenses (if publicly available)?: } Apache License 2.0 (for Tiramisu/LOOPer).
  \item {\bf Data licenses (if publicly available)?: } CC BY 4.0.
  \item {\bf Workflow automation framework used?: } Docker, Shell scripts.
  \item {\bf Archived (provide DOI)?: } Yes. Zenodo DOI: 10.5281/zenodo.16810084. URL: \url{https://doi.org/10.5281/zenodo.16810084}.
\end{itemize}
}

\subsection{Description}
\subsubsection{How to access}
The artifact is available on Zenodo (DOI: 10.5281/zenodo.16810084). The archive is a \texttt{.zip} file containing a \texttt{Dockerfile} and all necessary source code, scripts, and model weights required for the evaluation.

\subsubsection{Hardware dependencies}
A multi-core x86-64 CPU is required. The experiments in the paper were conducted on a dual-socket 12-core Intel Xeon E5-2695 v2 CPU. While the artifact will function on other modern x86-64 CPUs, the absolute performance results (e.g., execution times and speedups) are hardware-dependent and will vary.

\subsubsection{Software dependencies}
All software dependencies are encapsulated in the provided Docker container. The only requirement for the evaluator is a working Docker installation. The key components within the container are: Ubuntu 22.04 LTS, g++, CMake, Python 3.10, and PyTorch.

\subsubsection{Datasets}
The artifact uses three datasets:
\begin{enumerate}
    \item \textbf{PolyBench Implementation:} Included in the \texttt{PolyBench/} directory for running the performance evaluation.
    \item \textbf{Sample Training Data (80k points):} Included in the \texttt{cost\_model/} directory to demonstrate the functionality of the training pipeline.
    \item \textbf{Full Training Data (10M points):} This dataset was used to train the provided model. It is not included in the artifact due to its size but is publicly available here \url{https://huggingface.co/datasets/Mascinissa/LOOPerSet}.
\end{enumerate}

\subsubsection{Models}
The pre-trained weights for the LOOPer cost model, as evaluated in the paper, are included in the \texttt{cost-model-weights/} directory.

\subsection{Installation}
A working installation of Docker is the only prerequisite. The following commands build the container and start an interactive session.

\begin{enumerate}
    \item Build the Docker image from the artifact's root directory:
    \begin{verbatim}
docker build -t looper-pact25-ae .
    \end{verbatim}
    \item Run the container, mounting a local \texttt{results} directory to persist the output files:
{\footnotesize
    \begin{verbatim}
mkdir -p results
docker run --rm -it \
  -v $(pwd)/results:/mnt/LOOPer-pact25-ae/results \
  looper-pact25-ae /bin/bash
    \end{verbatim}}
\end{enumerate}
After these steps, the environment is fully configured and ready for experiments. All subsequent commands are to be run from inside the container.

\subsection{Experiment workflow}
The entire evaluation workflow is automated by the \texttt{run\_comparison.sh} script located in the \texttt{/mnt/LOOPer-pact25-ae/} directory inside the container. This script orchestrates the full comparison by:
\begin{enumerate}
    \item Calling \texttt{run\_LOOPer\_model.sh} to execute the model-guided LOOPer search on the PolyBench suite.
    \item Calling \texttt{run\_Pluto\_baseline.sh} to generate baseline performance using the Pluto compiler.
    \item (Optionally, for full evaluation) Calling \texttt{run\_LOOPer\_exec.sh} to run the measurement-guided search.
    \item Calling \texttt{aggregate\_results.py} to process the raw timing data, calculate geometric mean speedups, and print a final summary table to the console.
\end{enumerate}

\subsection{Evaluation and expected results}
The primary goal is to reproduce the performance comparison between LOOPer and Pluto, as shown in Figure 4.

\noindent\textbf{Main Evaluation (reproduces Figure 4 trend):}
To run the main evaluation on all 30 PolyBench benchmarks, execute the following command inside the container:
\begin{verbatim}
./run_comparison.sh basic 30
\end{verbatim}
This experiment takes approximately 3-4 hours. The expected output is a table printed to the console showing the geometric mean speedup of LOOPer over Pluto for each benchmark. The overall trend should confirm that LOOPer outperforms Pluto on a majority of benchmarks, consistent with the paper's findings. The final geometric mean across all benchmarks should be in a similar ballpark to the 1.42$\times$ reported in the paper, but the exact value will depend on the evaluation hardware.

\noindent\textbf{Quick ``Kick-the-tires'' Test:}
A fast validation run on 5 benchmarks can be performed to check the setup:
\begin{verbatim}
./run_comparison.sh basic 5
\end{verbatim}
This test should complete in approximately 30 minutes.

\subsection{Experiment customization}
The evaluation scripts are designed for easy customization. The list of benchmarks to run can be modified by editing the array at the top of the \texttt{run\_*.sh} scripts. Key hyperparameters of the LOOPer search, such as the beam size (\texttt{K}), can be configured in the header of \texttt{run\_LOOPer\_model.sh} and \texttt{run\_LOOPer\_exec.sh}.

\subsection{Notes}
For a more detailed, user-friendly guide with sample outputs, please refer to the \texttt{README.md} file included in the root of the artifact.

\subsection{Methodology}

Submission, reviewing and badging methodology:

\begin{itemize}
  \item \url{https://www.acm.org/publications/policies/artifact-review-and-badging-current}
  \item \url{https://cTuning.org/ae}
\end{itemize}

\end{document}